\documentclass[aps,prl,preprint,showkeys,groupedaddress]{revtex4-1}
\usepackage{graphicx}
\usepackage{subcaption}
\usepackage{relsize}
\usepackage{amsmath}
\usepackage{color}
\usepackage{tikz}
\usetikzlibrary{arrows,decorations}
\usepackage{booktabs}
\usepackage{amssymb}
\usepackage{dcolumn}
\usepackage{comment}
\usepackage{pgfplots}
\usepackage{color}
\pgfplotsset{compat=newest}
\usetikzlibrary{arrows,decorations,backgrounds,patterns,matrix,shapes,fit,calc,shadows,plotmarks,pgfplots.groupplots}
\usepackage{epstopdf}

\begin{document}


\title{Smoothing of the slowly extracted coasting beam from a synchrotron}


\author{R. Singh}
\email[Corresponding author: ]{r.singh@gsi.de}
\author{P. Forck}
\author{S. Sorge}
\affiliation{GSI Helmholtz Centre for Heavy Ion Research, Darmstadt, Germany}


\begin{abstract}
Slow extraction of beam from synchrotrons or storage rings as required by many fixed target experiments is performed by controlled excitation and feeding of a structural lattice resonance. Due to the sensitive nature of this resonant extraction process, temporal structure of the extracted beam is modulated by the minuscule current fluctuations present on the quadrupole magnet power supplies. Such a modulation lead to pile-ups in detectors and significant reduction in accumulated event statistics.  This contribution proposes and experimentally demonstrates that by introduction of further modulation on quadrupole currents with a specific amplitude and frequency,  the inherent power supply fluctuations are mitigated leading to a smoothening of the beam temporal structure. The slow extraction beam dynamics associated with this method are explained along with the operational results.

\end{abstract}


\maketitle


Controlled slow extraction of beam from synchrotrons and storage rings is required by a host of ``fixed target experiments" as well as hadron cancer therapy based on charged particle deposition on tumours~\cite{Amaldi}. The extracted beam is also referred to as ``spill". The extraction procedure is performed in two steps: a) exciting a lattice resonance and b) feeding the driven resonance by moving the betatron tunes of individual particle towards it. While most synchrotron facilities drive the horizontal third order resonance for slow extraction using sextupolar fields~\cite{pimms}, there exist a few variants of resonance feeding mechanisms (under operation) such as quadrupole driven~\cite{pimms}, betatron-core driven~\cite{pimms}, or RF knock-out~\cite{Hiramoto,Noda}. During the design phase, the choice of resonance feeding mechanisms is determined by extraction parameters such as beam energy, instantaneous momentum spread, total spill period and macro-shape of spill as requested by experiments as well as the ability to swiftly interrupt the extraction process~\cite{pimms,Noda}. The accelerator users typically assume that particle arrival times to the detectors is solely governed by Poisson statistics. However, with experience, most facilities and users have realized that the minuscule ripples and noise ($ < \Delta I/I \approx 10^{-5}$) on currents supplied to focusing magnets i.e. quadrupoles lead to significant temporal modulation of the extracted spills. These modulated spills are detrimental to statistics obtained by the experiments, where as much as 2/3rd of the delivered beam has been reported by experiments to be unusable~\cite{SEW}. Any improvements on the power supply side are either technologically unfeasible or would lead to long disruption in the operation of the facility.
Similarly, the hadron cancer therapy synchrotrons are also designed with an assumption of uniform spill delivery and reliability of the dose delivery is limited by non-uniformity of the spill and efforts are undergoing to improve the spill~\cite{Krantz}.
In order to deal with spill non-uniformities, the primary resonance feeding mechanisms in most facilities have been augmented by other techniques. One of first such techniques found in literature was proposed at CERN~\cite{VanDerMeer}, where the resonance feeding mechanism is assisted by a longitudinal stochastic noise, such that the particles are fed at a speed faster than the separatrix modulation induced by power supply noise. This "stochastic extraction" method reduced the effect of power supply ripples and is shown to work well for very long (in range of several minutes to hours) extraction times~\cite{Stockhorst,Hardt}. Other ideas also involved some sort of longitudinal gymnastics such as "rf phase displacement"~\cite{CappiSteinbach}, "rf channelling"~\cite{pimms} and bunched beam extraction~\cite{Forck} which all reduce the effect of power supply ripples at the cost of rf frequency modulation of the spill. 
Although these techniques reduce low frequency modulations on the spill, they actually hinder experiments which rely on smooth structure at time scales smaller than the revolution times. Spill smoothing based on feedback systems~\cite{Sato} is non-trivial to operate given that the slow extraction transfer function consists of a large (dynamic) delay  in the order of ms between particle extraction and its measurement in the spill. Feedback/servo systems have thus shown improvement only in low frequency or macro-spill regime ($< 50$ Hz)~\cite{HIT}  and limited success is reported in literature for micro-spill ($>50$ Hz)~\cite{JianShi}. 

In this report, 
we take a new view of the problem of uneven spill structure by primarily influencing the particle transit time, i.e. the time required by the particle to reach the electrostatic septum after it first becomes unstable. Consequently, the instantaneous variation of transit times for particles with different initial phase space co-ordinates and momenta being extracted at that same time instant referred to as "transit time spread" is also modified. It led us into developing a new method based on controlled tune modulation in correlation with the transit time referred to as "transit time dependent tune modulation". This method while smoothing the spill structure against power supply ripples, does not introduce any significant additional longitudinal structure at higher frequencies. The first application of this method is with the quadrupole driven resonance feeding mechanism, where the effect of power supply fluctuations on spill are expected to be the strongest~\cite{pimms}.
The recently concluded High Acceptance Di-Electron Spectrometer (HADES) experiment~\cite{hadesphase0} reported 50\% increase in recorded events compared to previous campaigns~\cite{SEW} as a direct result of these investigations.


\begin{figure}[htb]
\centering
\includegraphics[width=14cm]{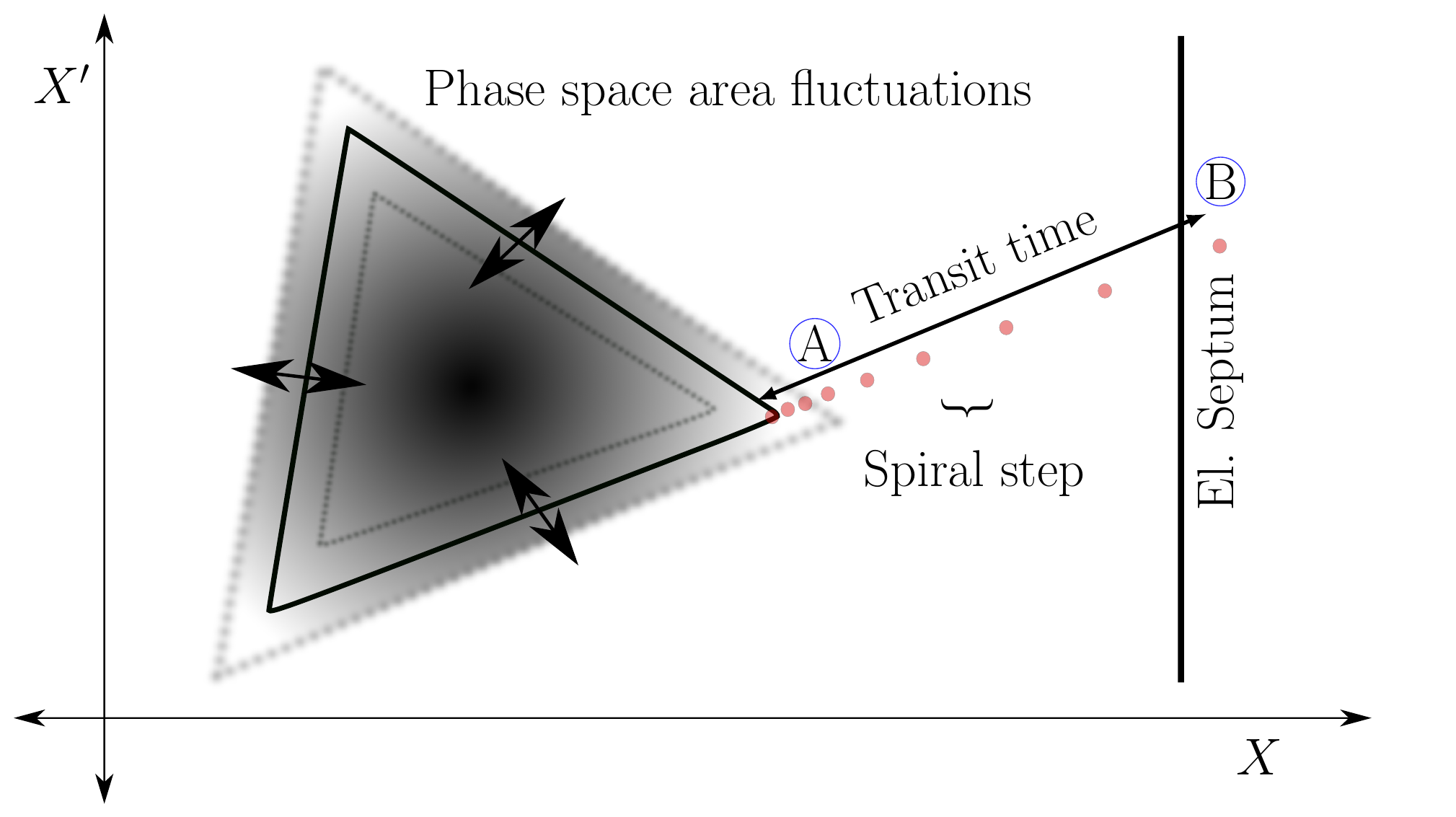}
\includegraphics[width=14cm]{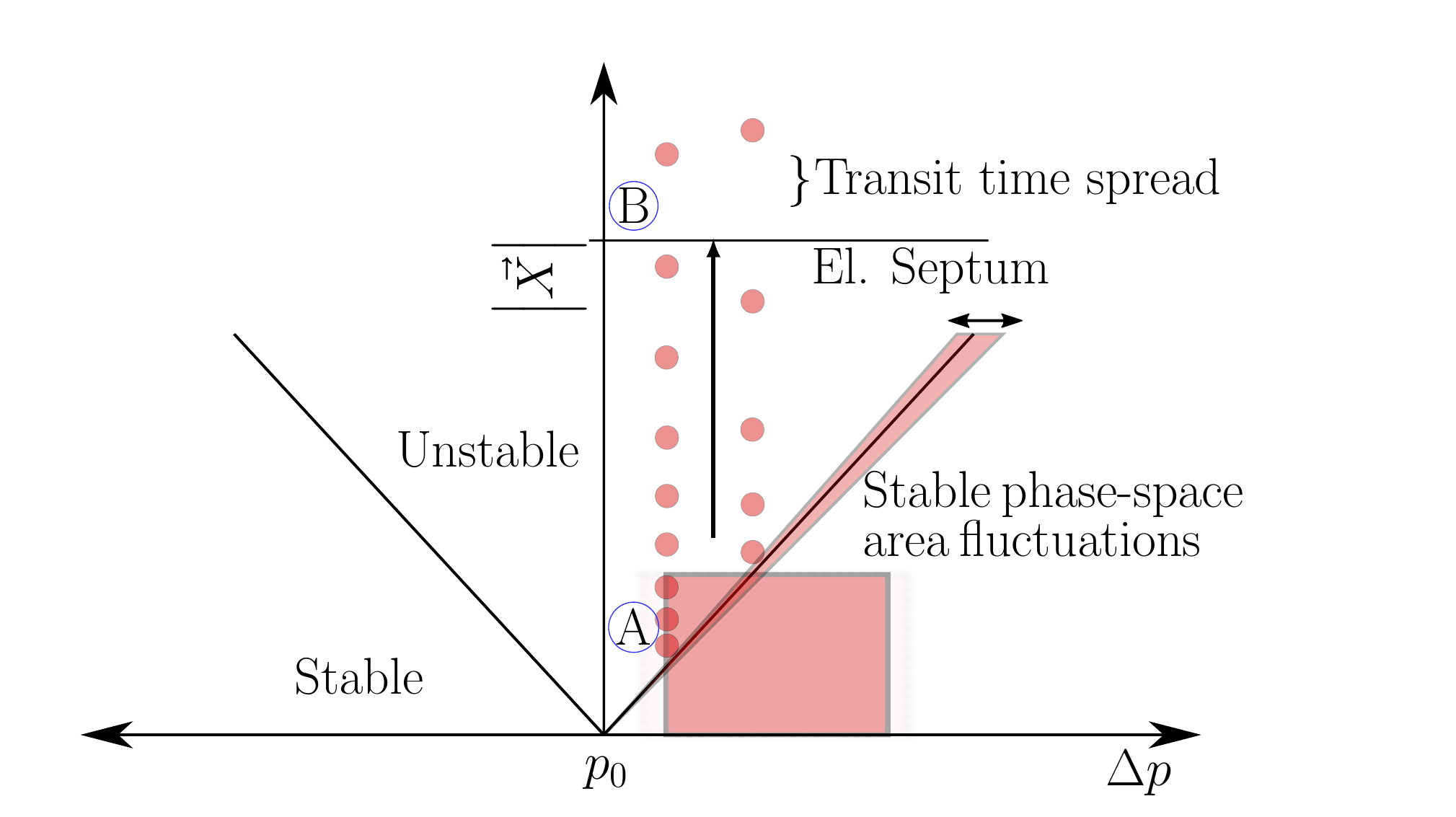}
\caption{(a) Schematic showing stable phase space area modulation due to power supply ripples for any specific tune and the particle transit towards the septum.(b) Steinbach diagram showing the spread in transit times due to amplitude and momentum distribution of fed particles.}
\label{fig:steinbach_phase_space}
\end{figure}

Proton Ion Medical Machine Study (PIMMS)~\cite{pimms} provides a comprehensive account of the slow extraction process. Here we discuss only the aspects relevant for this paper and try to utilize the PIMMS notations wherever possible. Figure~\ref{fig:steinbach_phase_space}(a) shows the stable phase space area for the considered one dimensional third order slow extraction mechanism. 
The triangular shape is a characteristic of third order resonance  described by Kobayashi theory \cite{kobayashi}. $\vec{X} \equiv (X,X^{'})$ is a phase space vector with the normalized 
coordinates 
 with
$X=\frac{x}{\sqrt{\beta_{x}}}$ and  $X^{'}=\sqrt{\beta_{x}} x^{'} + 
\frac{\alpha_{x}}{\sqrt{\beta_{x}}} x$. 
 where $\alpha_x,\beta_x$ are the Twiss parameters.
The size of phase space area is defined by unstable fixed points of 
betatron motion and is given as
\begin{equation} \label{eq_abs_xufp}
{A}_{stable} = \frac{4 \sqrt{3} \pi^{2}}{3} \left(\frac{\varepsilon_Q}{S_{v}} \right)^2, 
\end{equation} where 
$\varepsilon_Q = 6\pi(Q_{m}-Q_{r})$ is the difference between machine and
resonance tunes. $S_{v}$ is the strength of a virtual sextupole created by an arrangement of $N$ sextupoles governed  by the relation, $S_{v} \cdot {\rm e}^{3 {\rm i} \psi_{v}}=\sum \limits_{n} S_{n} {\rm e}^{3 {\rm i} \psi_{n}}
$
with the normalised sextupole strength of the $n^{\rm{th}}$  sextupole $S_{n}= \frac{1}{2} \beta_{x,n}^{3/2} (k_{2} L)_{n}$
and the phase advance $\psi_{n}$ between the considered location which is 
usually that of the electrostatic septum and the location of the $n^{\rm{th}}$ 
sextupole, where $\psi_{n}$ corresponds to the third integer tune of the 
resonance $Q_{r}$ and $\psi_{v}$ determines the orientation of the stable phase space. The Steinbach diagram in Fig.~\ref{fig:steinbach_phase_space}(b) shows the separation of stable and unstable area as a function of particle momentum $\Delta p$. The slope of separating line is, $\left| \vec{X} \right| /\Delta p \propto (Q_r\xi)/(p_0S_v)$ where $\xi$ is the chromaticity in the plane of extraction and $p_0$ is the momentum of zero amplitude particle at resonance tune $Q_r$.
In quadrupole driven extraction, the stable phase area is slowly shrunk with time by moving the machine tune  $Q_{m}$ towards resonance $Q_{r}$ such that the particles (starting from larger amplitudes/action) leave the stable area and traverse towards the extraction septum. Upon crossing the electrostatic septum, particles obtain a kick such that the particle trajectory passes through another magnetic kicker, thus spilling it out of the synchrotron.
 For a particle with a certain amplitude $\left| \vec{X} \right|$, the transit time is primarily determined by the distance to resonance $\varepsilon_{Q}$ at which the particle becomes unstable, which in turn is directly proportional to the strength of resonance ($S_{v}$) as shown by Eq. 4.17 in~\cite{pimms}
\begin{eqnarray}\label{eq:transit_time}
T_{tr}&\propto&\frac{1}{\varepsilon_Q}
\end{eqnarray} 
For a particle beam with a given emittance and finite momentum spread, there is corresponding spread in transit times. This is also schematically shown in  Fig.~\ref{fig:steinbach_phase_space}(b). The transit time spread scales with the "mean" transit time $\overline{T_{tr}}$,
 \begin{eqnarray}\label{eq:transit_time_spread}
 \overline{T_{tr}}&\propto&\frac{1}{\overline{\varepsilon_Q}} \\
 \Delta T_{tr} &=& \frac{d T_{tr}}{d\varepsilon_Q} \Delta \varepsilon_Q \propto \frac{1}{\overline{\varepsilon_Q}^2} \Delta \varepsilon_Q
 \end{eqnarray} as discussed in~\cite{SorgeIOP}. 
The spiral step growth at any time step is given as a function of sextupole field strength and coordinates in the previous time step $(X_0,X_{0}')$ as $\Delta X_1 \propto S_v {X_0'}(3^{1/4}\sqrt{A_{stable}}/2+1.5X_0)$ .

The fluctuations of $A_{stable}$ shown in Fig.~\ref{fig:steinbach_phase_space} are dominated by the field ripples of quadrupole and sextupole magnets.
The modulations in size of the stable phase space area for independent quadrupole and sextupole ripples  is estimated as,
\begin{eqnarray}
\delta A_{stable}
&=&  2A_{stable}  \left( 6\pi\left| 
\frac{\delta Q_{m}}{\varepsilon_Q} \right| + \left| \frac{\delta S_{v}}{S_{v}} 
\right| \right) 
\end{eqnarray}
$\delta \varepsilon_Q=6\pi\delta Q_{m}$ because $Q_{r}={\rm const}$. If we assume that quadrupole and sextupole fluctuations are of the same order, i.e. $\left| \delta Q_{m}/Q_{m} \right| \sim \left| \delta S_{v}/S_{v} \right|$,
the contribution of the quadrupole ripples has a larger influence since
$|\varepsilon_Q| \ll Q_{m}$ so that 
$\left| \delta Q_{m}/\varepsilon_Q \right| \gg \left| \delta Q_{m}/Q_{m} \right|$ 
resulting in 
$\left|\delta Q_{m}/\varepsilon_Q \right| \gg \left| 
\delta S_{v}/S_{v} \right|
$ as also observed at GSI SIS-18 and CERN SPS~\cite{Prieto}.

\begin{figure}[htb]
\centering
\includegraphics[width=10cm]{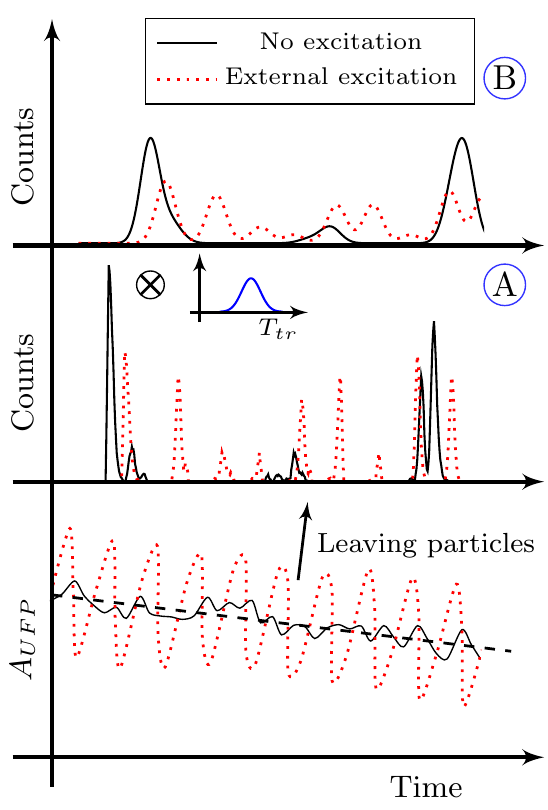}
\caption{Bottom: Stable phase space area modulation due to inherent power supply ripples (black solid lines) and when external ripples are introduced on quadrupoles (red dotted lines). Middle: Spill just after crossing the separatrix, Point A in Fig.~\ref{fig:steinbach_phase_space}. Top: Spill at the measurement location, Point B in Fig.~\ref{fig:steinbach_phase_space} after convolution with transit time distribution.}
\label{fig:explain_effect}
\end{figure}
The effect of fluctuations on the measured spill structure is illustrated in Fig.~\ref{fig:explain_effect}. The dashed line in Fig.~\ref{fig:explain_effect} (bottom) show the controlled shrinkage of stable phase space area as it is performed with time, while the solid black line depicts the undesired modulation $\delta A_{stable}$ due to quadrupole field fluctuations for particles with a specific momentum. The nature of fluctuations is similar irrespective of particle momentum although the absolute stable phase space area is different for different momenta as indicated in Fig.~\ref{fig:steinbach_phase_space}(b). As the stable area shrinks, particles are "released" and their relative counts are shown in Figure~\ref{fig:explain_effect} (middle) shortly after crossing their respective momentum dependent separatrices denoted by point $A$ (also shown in Figure~\ref{fig:steinbach_phase_space}). Following that, the released particle are convolved with the instantaneous transit time distribution $\Delta T_{tr}$ while traversing from point $A$ to the Point $B$ (top) which can be seen as the point of "spill measurement or user experiment".The convolution has the effect of low pass filtering and higher frequency components above a certain cut-off frequency $f_{cut} \propto \Delta T_{tr}$ are suppressed.
 
\begin{figure}[htb]
\centering
\includegraphics[width=8 cm]{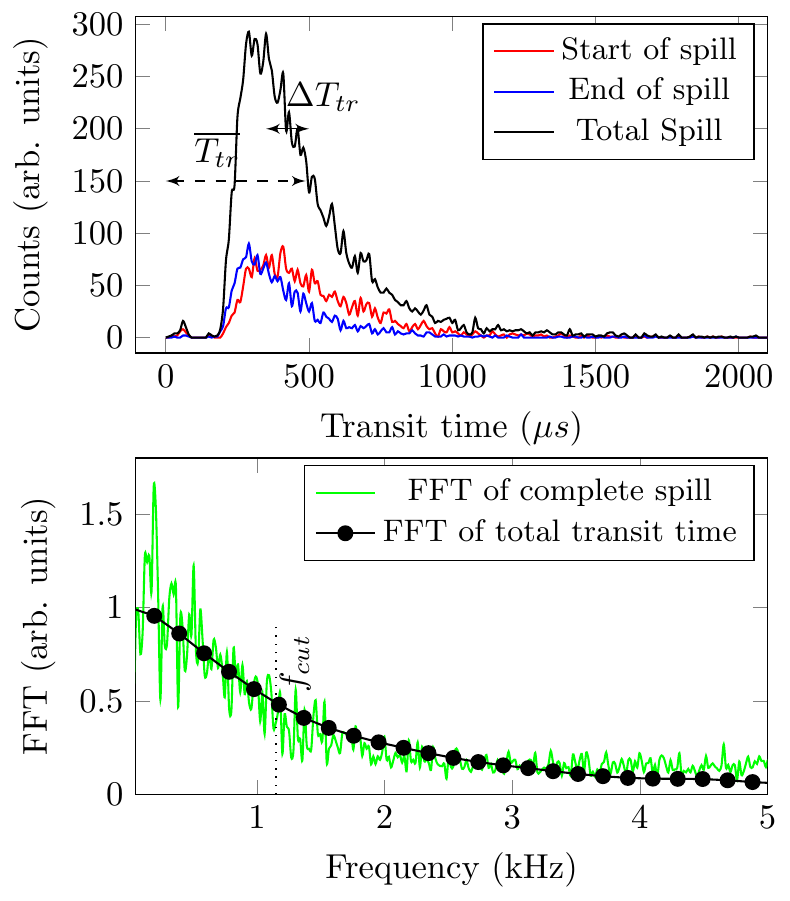}
\caption{Transit time distribution and the comparison of the Fourier transform with the spill spectra.}
\label{fig:sim_transit_spectrum}
\end{figure}

The transit time distribution parameters i.e. $\overline{T_{tr}}$ and $\Delta T_{tr}$ primarily depend on the slow extraction settings which in turn are determined by the beam parameters, beam emittance, beam energy and its spread, as well as length of spill discussed as elaborated in ~\cite{SorgeIOP}. An imprint of the transit time spread $\Delta T_{tr}$ can be seen on the spill spectrum by means of the cut-off frequency $f_{cut}$, while $\overline{T_{tr}}$ is expected to scale with $\Delta T_{tr}$ in accordance to Eq.~ \ref{eq:transit_time_spread}. Figure~\ref{fig:sim_transit_spectrum} shows the evaluated transit time distributions from particle tracking simulations during a typical spill for SIS-18 synchrotron. The simulated beam energy is 300 MeV/u $C^{6+}$ beam with mean extraction rate of $2\times 10^5$/s in 0.5 s. The virtual sextupole setting $S_v= 5.5$ $m^{-1/2}$ and the tune $Q_x$ is driven from 4.327 into the third order resonance $4.\overline{33}$. The revolution frequency is $1.1$ $\mu$s and momentum spread $\Delta p /p_0 =5 \cdot 10^{-4} (2\sigma)$ and natural chromaticity $\xi = -0.94$. The distribution mean $\overline{T_{tr}}$ and spread $ \Delta T_{tr}$ evolve and are correlated $ \overline{T_{tr}} \propto \Delta T_{tr}$ at all instances during the spill (blue and red curves in Fig.~\ref{fig:sim_transit_spectrum}). From hereon, we will utilize for the "total" transit time distribution for the discussion (black curve in Fig.~\ref{fig:sim_transit_spectrum}) unless otherwise specified. The mean transit time $\overline{T_{tr}}$ is $488$ $\mu$s. The filtering effect of transit time spread is shown in Fig.~\ref{fig:sim_transit_spectrum} (bottom), where the Fourier transform of the transit time distribution is compared with the respective spill spectrum. Since the "total" transit time distribution resembles a decaying exponential, the $\Delta T_{tr}$, can be recovered from the frequency corresponding to half maximum in its Fourier transform i.e. $\Delta T_{tr}=1/(2 \pi f_{cut})=138$ $\mu s$.

\begin{figure}[htb]
\centering
\includegraphics[width=8.2cm]{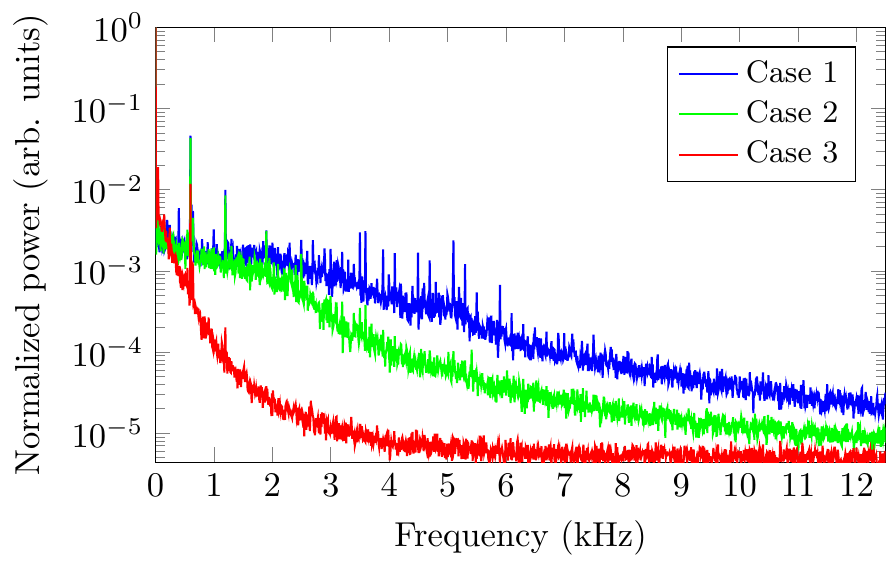}
\includegraphics[width=8.2cm]{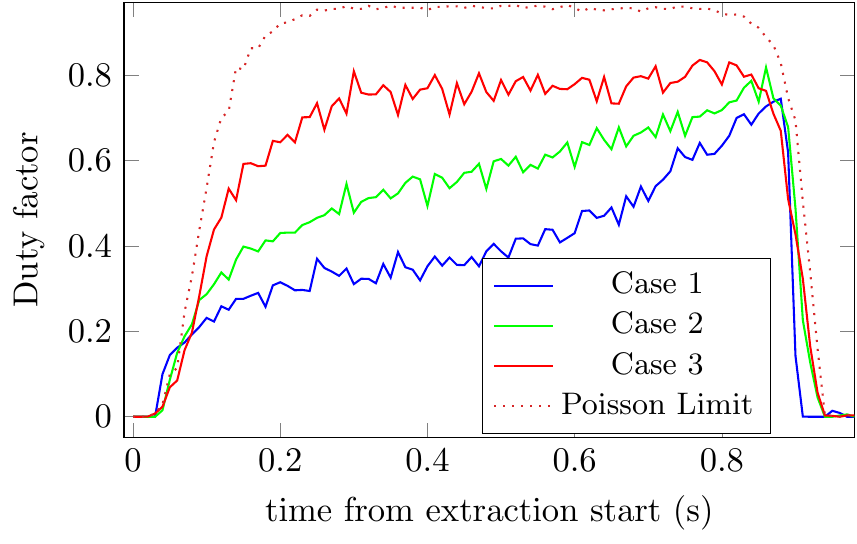}
\includegraphics[width=8.2cm]{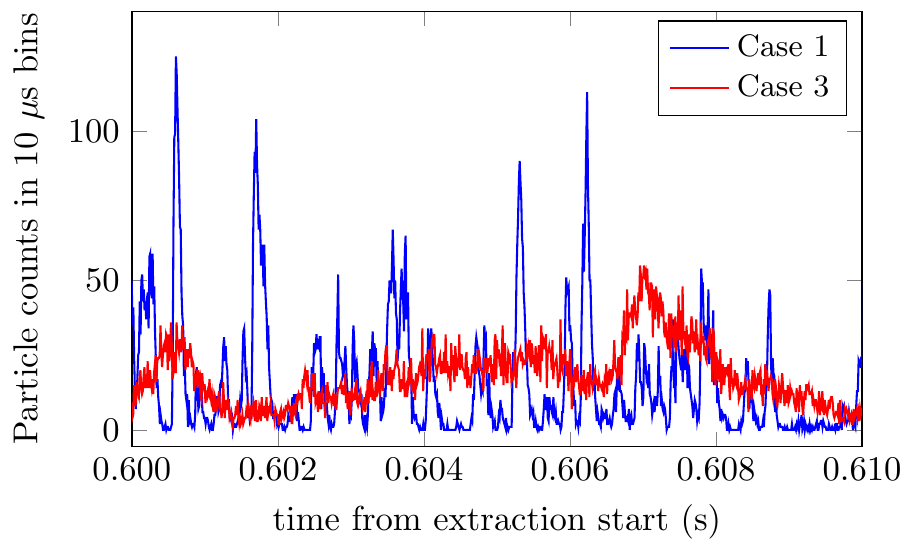}
\caption{Comparison of beam spill (bottom), its duty factor (middle) and the power spectra (top) for case (1) stronger sextupole strength and large emittance, case (2) weaker sextupole strength and large emittance and case (3) weaker sextupole strength and small emittance.}
\label{fig:beam_size_effect}
\end{figure}

Our first approach towards spill smoothening looked at influencing the transit time spread. As evident from Eq.~\ref{eq:transit_time} and Fig.~\ref{fig:steinbach_phase_space}(b), the transit time spread $\Delta T_{tr}$ can be increased by reducing the initial $\overline{\varepsilon_Q}$ which in turn is performed either by reducing the sextupole fields $S_v$~\cite{SorgeIOP,SinghIOP} or reducing the beam emittance. This is  demonstrated by comparing three scenarios with a 300 MeV/u $C^{6+}$ beam with mean rate of $2\times 10^6$/s. Case (1) is that of a "typical" working point in terms of sextupole strength $S_v = 11.0$ $m^{-1/2}$ optimized for minimizing beam losses and nominal beam emittance of 15 mm-mrad , second (case 2) is that of reduced sextupole strength $S_v = 5.5$ $m^{-1/2}$ for nominal injection beam emittance of 15 mm-mrad, and the third (case 3) is for a small emittance of 4 mm-mrad at the lower sextupole setting $S_v = 5.5$ $m^{-1/2}$. These are compared against their respective power spectra in Fig.~\ref{fig:beam_size_effect}. It is clear that, for the last case, the $f_{cut}$ is the lowest as the higher frequency fluctuations are curbed and results in the best (smooth) micro-spill structure. Another important observation is that the transit time spread increases as $\overline{\varepsilon_Q} $ approaches zero towards the end of spill, and thus the duty factor or quality of spill is improved. The ``duty factor"~\cite{pimms,SinghIOP} $F$ is defined as the ratio $F=\langle N_p \rangle^2/\langle N_p^2 \rangle$ where $N_p$ is the number of particle counts per measurement time $T_m = 10$ $\mu$s and $\langle \cdot \rangle$ represents the operation "mean". Each duty factor is calculated over 1000 such measurements if the characterization bin is $T_c =10$ ms i.e $p = 1 \rightarrow 1000$. Notably, some studies took an exactly opposite strategy of increasing the sextupole fields $S_v$ and distance to resonance $\varepsilon_Q$ for micro-spill improvement~\cite{Mizushima}. The limitation of our first approach is that both the sextupole strength and beam size cannot be arbitrarily reduced without effective loss is statistics for the experiments. As discussed earlier, lower sextupole strength results in a smaller spiral step $\Delta X$, which below a certain threshold leads to losses on the septum wires.
Similarly, reducing the beam size at multi-turn injection synchrotrons typically results in lower intensities, and thus effective loss in event rate for experiments. Transverse beam cooling also needs significant time and thus is not a viable option for experiments relying on large cumulative statistics. Therefore, a new method independent of sextupole strength and beam emittance was developed and is discussed next.

The foundation of this method is an application of external current modulation with amplitudes larger than the inherent ripples into the quadrupole power supplies with a scheme discussed in~\cite{SinghIOP}. This leads to a modulation of machine tune $Q_m$ and thus the stable space area as marked with dotted red lines in Fig.~\ref{fig:explain_effect} (bottom).
 The spill smoothing works as follows; due to relatively large tune modulation at the introduced frequency compared to inherent ripples, particles undergo a forced release from the stable phase space area at the applied frequency (Fig.~\ref{fig:explain_effect} (middle)). The introduced modulation prevents the lower frequency inherent ripples from "feeding" on the particles, which is in principle similar to a fast separatrix crossing in the $rf$ based methods~\cite{CappiSteinbach} and thus low frequency inherent ripples are strongly suppressed in spill.
In order to ensure that the spill does not form a significant modulation at the introduced frequency,i.e. it has to be higher than the cut-off frequency $f_{ext} > f_{cut}$.
The upper limit on the introduced frequency is given by the condition that the particles are not re-captured in the other half of the cycle. This conditions hint at a certain scaling of introduced modulation frequency with the mean transit time $f_{ext} \propto 1/\overline{T_{tr}}$ and therefore $f_{cut}$.

 \begin{figure}[htb]
\centering
\includegraphics[width=7 cm]{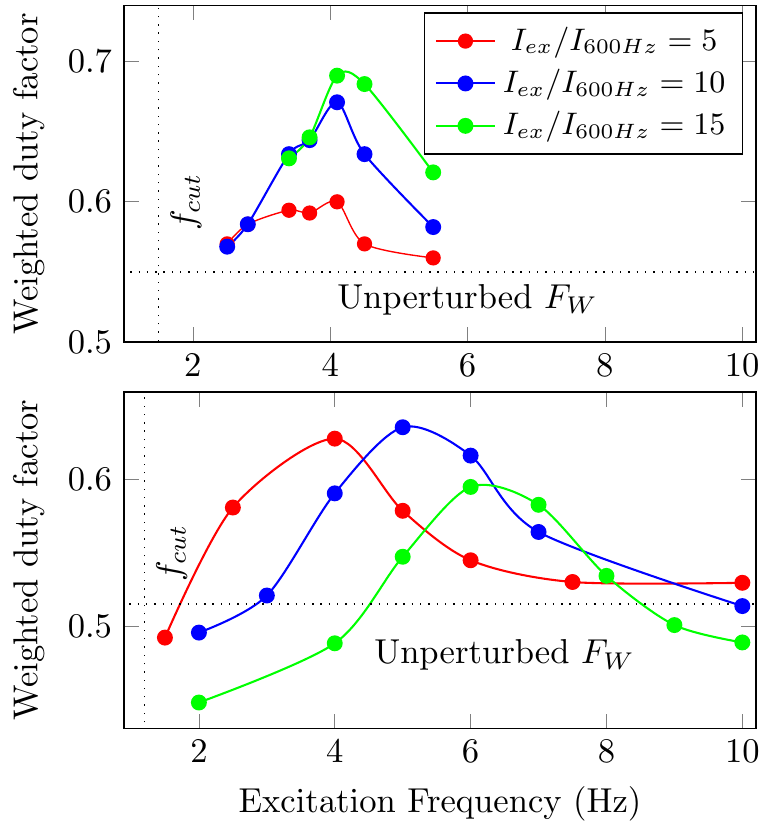}
\includegraphics[width=7 cm]{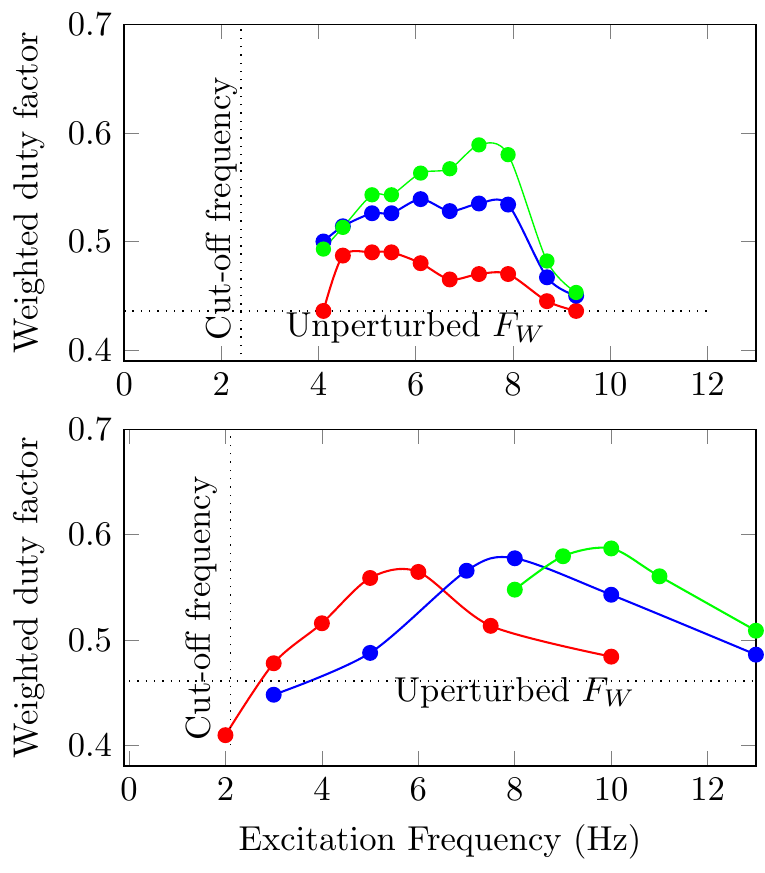}
\caption{Weighted duty factor in experiment (top) and simulations (bottom) as a function of excitation frequency and amplitude. The left plots are with $S_v = 5.5$ $m^{-1/2}$ and right plots are for $S_v = 11$ $m^{-1/2}$.}
\label{fig:Sim1600}
\end{figure}

Figure~\ref{fig:Sim1600} show the simulation and experimental results of the dependence of spill quality on the excitation amplitude and frequency. The legend marks the set amplitude with respect to the inherent 600 Hz ripple amplitude. During experiments, the introduced excitation was distorted as a function of frequency for $I_{ex}/I_{600 Hz} = 15$ case, and might have resulted in lower effective amplitude. The weighted duty factor $F_W $ is the weighted sum of duty factors calculated along the spill for each $k^{th}$ characterization bin ($T_c$=10 ms) resulting in a single number per excitation frequency and amplitude setting.
\begin{equation}
F_W= \sum_{i=1}^N \frac{F_k \cdot \langle N \rangle_k}{\langle N \rangle_k}
\end{equation}
The extraction parameters are the same as discussed earlier both for simulations and experiments, the only difference is in the extraction rate which is a factor 10 higher in experiments i.e. $10^6/s$. There are two clear trends, that the optimal excitation frequency scales with the cut-off frequency i.e. $\approx 4$ times of the cut-off frequency  and there is a dependence between amplitude and frequency of the introduced modulation.
The first observation establishes the relation between transit time spread and optimal frequency i.e $f_{ext} = (3-5) \cdot f_{cut}$ and utilizing example case of the simulation where, $\Delta T_{tr}/\overline{T_{tr}}= 138/488 = 0.28$, one arrives at the useful frequency range $f_{cut} < f_{ext}  < (12-20)/\overline{T_{tr}})$. The amplitude of excitation should be significantly higher than the inherent ripple amplitude such that it forces the particle release ahead of inherent ripples. On the other hand, it should be low enough such that particles released in the previous cycle are not re-captured (the same argument as for high frequency modulation) and the total length of spill is not affected due to associated tune modulation. The optimal value in simulation is a factor of 10 times higher than the inherent ripple amplitude which corresponds to $2\%$ of total current change during the slow extraction tune ramp in our experiment. The amplitude and frequency dependence reported in Fig.~\ref{fig:Sim1600} is related with dependence between spiral step and transit time, i.e. for smaller transit times, the spiral step is larger and thus a larger amplitude excitation can be employed without recapturing particles. 
 At much higher frequencies $f_{ext} \gg 1/\overline{T_{tr}}$, most particles would not be able to follow the introduced oscillations of stable area and no periodic release of particles at the introduced frequency will occur and thus spill quality is not affected by external modulation.  
Further, in a quadrupole driven extraction process, where $\overline{T_{tr}}$ varies during the extraction, the exciting frequency should follow it during the extraction.

\begin{figure}[htb]
\centering
\includegraphics[width=8.2 cm]{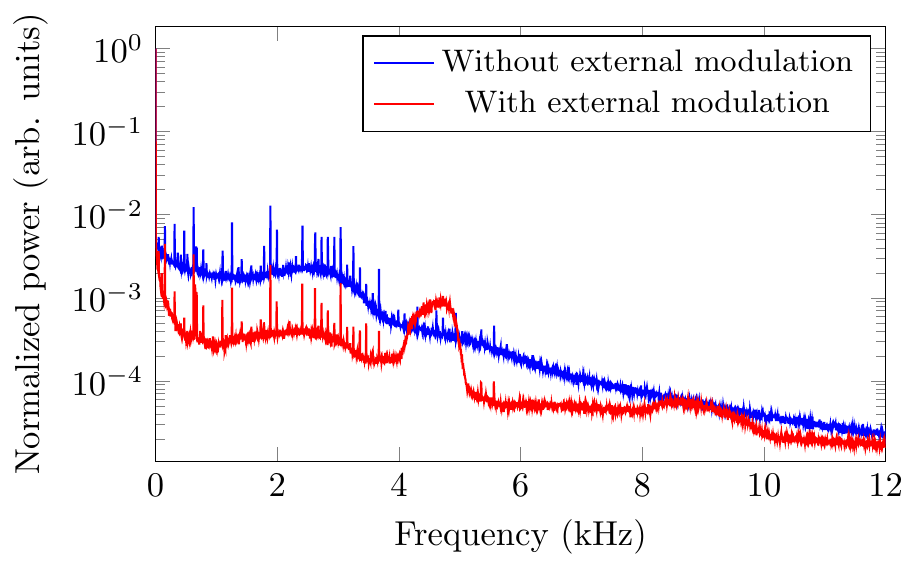}
\includegraphics[width=8.2 cm]{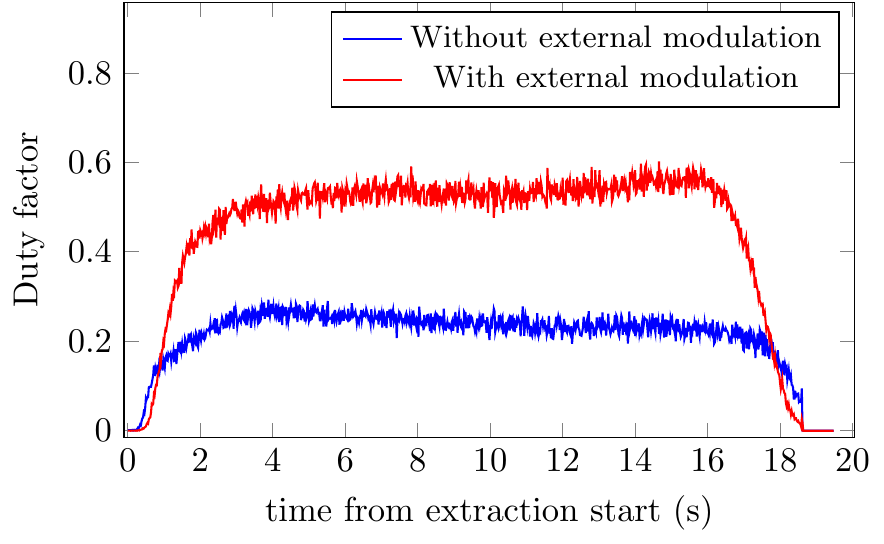}
\includegraphics[width=8.2 cm]{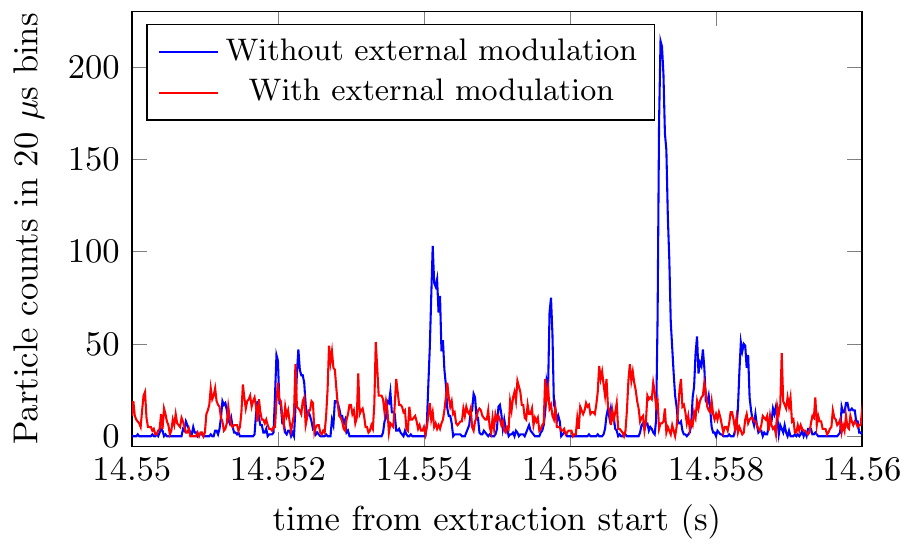}
\caption{Comparison of beam spill (bottom), its duty factor (middle) and the power spectra (top) with and without external modulation of tune.}
\label{fig:external_ripples}
\end{figure}
\begin{figure}[htb]
\centering
\includegraphics[width=17 cm]{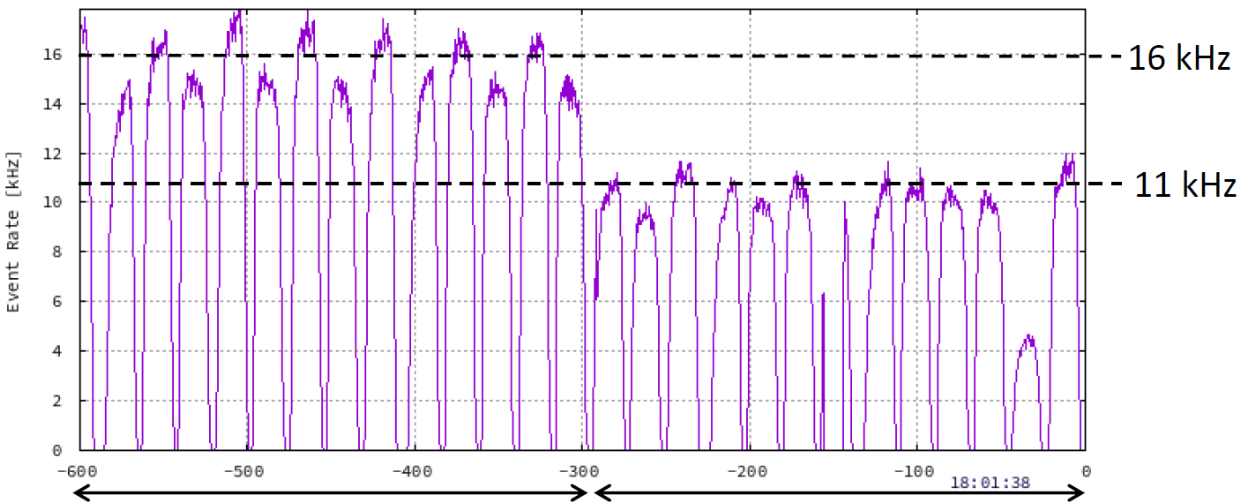}
\caption{The event rate registered in the HADES detector with tune modulation (from 300 to 600 s) and no modulation (0 - 300 s). The x-axis shows elapsed time in seconds with respect to the last measurement. Each extraction spill lasted for 20 s.(Courtesy: J. Pietraszko and M. Traxler from the HADES collaboration).}
\label{fig:HADES}
\end{figure}

Figure~\ref{fig:external_ripples} shows the spill spectra and duty factors with and without tune modulation for a high energy beam of 1.58 GeV/u $Ag^{45+}$ corresponding to the highest rigidity at SIS-18 for the HADES experiment~\cite{hadesphase0}. In this case, the duty factor does not improve from start to end of extraction (unlike shown in Fig~\ref{fig:beam_size_effect}). The reason for that is the insufficient voltage on the electrostatic septum at the highest SIS-18 rigidity. Insufficient kick from the septum leads to loss of particles with smaller spiral step $\Delta X$ (or larger transit time) and thus the effective transit time spread distribution $\Delta T_{tr}$ remains roughly the same for the whole extraction. In Fig.~\ref{fig:external_ripples} (top), the wiggly shape of spectrum in the "passband" is due to transit time dependent septum losses.
This also results in a spill quality of $F\approx 0.2$ which is worse than the ones obtained at lower energies. 
A sweep of excitation frequency from $5$ kHz to $3.8$ kHz was introduced in order to account for the transit time variation during extraction. As a result, all the lower frequency components are suppressed by $\approx$ 10 dB in the power spectra shown in Fig.~\ref{fig:external_ripples} (top). This led to a factor 2.5 improvement in the duty factor and a much smoother spill as shown in Fig.~\ref{fig:external_ripples} (middle and down respectively). The HADES experiment reported an increase by a factor of 1.5 in cumulative statistics. The corresponding event rates (courtesy of HADES collaboration) are shown in Fig.~\ref{fig:HADES}. The smoother spill structure is further expected to reduce the number of discarded events during post processing and increase the effective event rate useful for the experiment.

The transit time dependent tune modulation method for spill smoothing is described in this report. It is operationally straightforward and immediately renders itself to other facilities and experiments facing issues with micro-spill non-uniformity. In summary, the steps to be followed are; an estimate of the cut-off frequencies at the start and end of spill is made from the spill spectrum. A frequency sweep between the two optimal frequencies satisfying the criterion $f_{cut }< f_{ext} < (3-5) \cdot f_{cut}$ with an amplitude of $1-2\%$ of the quadrupole current change due to tune ramp during extraction is introduced. Since the full transit time distribution is not accessible during experiments, some empirical tweaking of amplitude and frequency should be performed to obtain the optimal smoothing. As a next step, waveforms other than a simple sinusoid for the introduced tune modulation will be studied. On a longer term, particle amplitude dependent external tune modulation using non-linear fields could be beneficial.


\begin{acknowledgments}
Horst Welker and Andrzej Stafiniak are gratefully acknowledged for their consistent support in introducing the ripples in quadrupoles power supplies. SIS-18 team provided support in bringing the technique into regular operation on a very short notice.
\end{acknowledgments}




\end{document}